\documentclass[prb,preprint]{revtex4-1} 
\usepackage[margin=2.5cm,footskip=0.25in]{geometry}
\usepackage[english]{babel}
\usepackage[utf8]{inputenc}
\usepackage{graphicx}
\usepackage{subcaption}
\usepackage{caption}
\usepackage{amsmath}  
\usepackage{amsfonts} 
\usepackage{amssymb}
\usepackage{xcolor,soul}
\usepackage{physics}
\usepackage{setspace}

\begin{document}
\def\bibsection{\section*{\refname}} 


\title{Dynamics of falling raindrops}
\author{Yashvir Tibrewal }
\affiliation{NPS International School, Singapore}
\email{yashvirtib@gmail.com}
\author{Nishchal Dwivedi}
\affiliation{Department of Basic Science and Humanities, SVKM’s NMIMS Mukesh Patel School of Technology Management \& Engineering, Mumbai, India }
\email{nishchal.dwivedi@nmims.edu}

\begin{abstract}

Studying water droplets is a rich lesson in fields of fluid dynamics, nonlinear systems, and differential equations. Understanding various physical aspects of raindrops can help us in understanding drop dynamics, rainfall density estimation, size distributions which can be grant insights in the fields of meteorology, hydrology, and climate science. This work identifies the real world significance in developing more accurate atmospheric attenuation correction algorithms that could potentially overcome the scattering effect of rain on radio and micro wave communication. This works presents and aims to compile some historical work which focuses on the influence of surface tension in the droplet nucleation and formation, raindrop oscillation, burst cycles, and transformation into parachute-like shapes before fragmentation. The interplay between surface tension and air resistance in raindrop dynamics is also explored. 
 
\end{abstract}
\maketitle

\section{Introduction}

Rainfall forms a fundamental component of Earth's hydrological cycle and plays a vital role in shaping our climate and environment \cite{carvalho2014prediction}. The exact intensity of rainfall is significantly affected by size distributions  shape variations of individual droplets\cite{seliga1981preliminary}. By understanding the driving forces behind their shapes it would be easier to classify the falling raindrops accurately. This work explores the exact dynamics and distribution of forces that cause the topological changes observed as rainfalls descend. 

Studying raindrop dynamics is also of practical interest due to ability of rain to scatter radio and microwave signals. In a process known as atmospheric attenuation raindrops scatter the incoming electromagnetic signal thereby degrading it. Therefore studying this process can help improve all technology which rely on radio and micro wave communication including wireless communication systems, radar technologies, satellite communications, and remote sensing devices\cite{thurai2014investigating}. 

By modeling the transformations of falling raindrops we will be able to probe this scattering effect on electromagnetic radiation potentially developing more advanced algorithms to correct or adjust for this degradation in signal. The mitigation of impact atmospheric attenuation of the transmission and reception of signals would revoluntise several fields. It could improve global wireless networks of communications. When the effects of changes in raindrop form are taken into consideration, radar systems can benefit from more precise readings and enhanced target recognition skills. Through the use of accurate attenuation correction techniques, satellite communications may increase data transmission speeds and improve link availability. Furthermore, remote sensing will be able to more intelligently gather more accurate data of atmospheric parameters and climatic conditions on Earth\cite{wei2021remote}. Overall this would greatly improve data transmission, communication and reception advancing several fields.

The study and modeling of raindrop oscillation have encountered considerable challenges due to the intricate nature of rain phenomena. Raindrops exhibit a burst cycle, especially those with a Weber number exceeding 6. The process commences with raindrop formation, where droplets cluster around nucleation sites within clouds, such as minute dust particles. Initially, the raindrop assumes a spherical configuration to minimize surface tension. However, as it descends, it experiences progressively greater air resistance, resulting in one side flattening. Over time, this flattened side gradually becomes more concave, ultimately transforming the raindrop into a parachute-like shape with a toroidal rim. Eventually, the air resistance becomes overpowering, causing the rim to become unstable, leading to the raindrop bursting into smaller fragments—a phenomenon commonly referred to as bag break-up.
Notably, as long as the Weber number (We) remains moderate, satisfying the condition We < 6, the radius of the raindrop consistently oscillates around a mean value. This paper aims to provide a summary of previous studies conducted in this area \cite{seliga1981preliminary,tokay1996field,kostinski2009raindrops}. Expanding on their works by considering the influence of surface tension on drop formation and later effects in the fragmentation cycle.

\section{Aim}
This research considers previous investagation carried out to investigate raindrop shape transformations and size distributions. Then outlines the mathematical parameters which describe the nucleation of rain droplets reasoning their formation. It describes raindrop size distribution as the intensity of the rainfall varies. 

\section{Previous Experiments}
Throughout history have been several incredibly designed experiments that overcome technological limitations and find novel ways to record the oscillations and size distributions of raindrops. 

\subsection{Bentley and Lenard}

The data for this distributions has been collected from as far back as 1904 in the experiments by Bentley and Lenard \cite{wa1904studies,blanchard1972bentley}. Wilson A. Bentley and Phillip Lenard, two pioneering scientists in cloud physics, conducted groundbreaking research on raindrop formation in 1904. 

Bentley collected numerous raindrop samples by allowing them to fall into a layer of fine, uncompact flour within a shallow tin receptacle. After letting the raindrops stay until the resulting dough pellets dried and hardened, he meticulously measured and extracted them, closely matching the original raindrop sizes. He then calibrated his results by taking water drops on known size and dropping the from known heights to measure the size of the flour pellets produced allowing him to categorise the size distribution of raindrops. During his observations, Bentley recorded crucial meteorological data, including temperature, cloud height, storm type, and more, for each raindrop sample. Through this method, he managed to obtain 314 sets of raindrop impressions from 70 different storms. Bentley's findings unveiled striking variations in raindrop dimensions, ranging from large drops with 1/4 inch in diameter to tiny ones measuring less than 1/500 inch. He organized the data into tabular form, illustrating the relative frequency of various raindrop sizes in the collected samples. Furthermore, Bentley examined the distribution of raindrops within different segments of storms, discovering that larger raindrops were more prevalent in the extra-central, western, or receding storm segments. He also investigated the impact of various cloud types on raindrop size, noting that clouds with significant vertical dimensions tended to produce larger raindrops. 

Bentley's method of raindrop observation involved letting raindrops fall into a layer of fine flour in a shallow tin receptacle. He allowed the raindrops to remain in the flour until the resulting dough pellets dried and hardened, closely matching the original raindrop sizes. With meticulous data collection and meteorological records, Bentley obtained an 344 drop-size distributions from 70 different storms. His observations revealed the variation in raindrop sizes based on cloud types, with clouds of significant vertical dimensions producing larger raindrops.

On the other hand, Phillip Lenard's methodology for studying raindrop dynamics involved the use of a vertical wind tunnel, which he designed as a prototype for the wind tunnels widely used today. In this setup, he adjusted the upward speed of the air flow to match the fall speed of waterdrops, effectively suspending the drops relative to an observer. While he could only suspend the drops for a few seconds, it was enough to determine the fall speed of waterdrops as a function of size. Lenard's experiments revealed that the fall speed of raindrops increased with size up to about 4.5-mm diameter, beyond which it plateaued due to the change in the shape of the falling drops.

Bentleys and Lenards work may not have garnered the recognition it deserved during their time. Their ingenious experiments and meticulous measures of data collection set a new standard for studying and making observations about the rain. Nevertheless their research laid the foundation for modern cloud physics and continues to influence emerging ideas and theories in the field. 

\subsection{Dual-polarization differential-reflectivity (Z-DR) radar technique }
The research conducted by Seliga, Bringi, and others \cite{seliga1981preliminary} focused on improving the accuracy of measuring rainfall rates and precipitation parameters using the differential reflectivity (Z-DR) technique. This method employed two simultaneous remote measurables, ZH (reflectivity factor), and Z-DR (differential reflectivity), to account for variations in the raindrop size distribution (DSD) parameters, which conventional methods could not\cite{direskeneli1983differential}.

- ZH (Reflectivity): Provided insights into the intensity of precipitation in a particular area and measured the power returned to the radar receiver.
- Z-DR (Differential Reflectivity): Used to distinguish between the types of precipitation such as snow,hail,rain and ice pellets using parameters relating to size and shape

The experiments carried out were primarily to test the limitations of the Z-DR technique in various environment conditions. Data was collected using disdrometers and radar measurements in various rainfall events. Aftering anaylizing the results and comparing the it with the rates determined Z-DR method researchers found that  Z-DR method signgifcantly reduced the scatter in relation to conventional methods of rainfall analysis. Furthermore they exploring how to further their experiment through empirical relations between test parameters by considering gamma distribution possibling increasing the overall accuracy.
The researchers investigated the effects of variations in DSD shape, uncertainties in drop size measurements, and other factors on the accuracy of the Z-DR method. Their designed experiments involved collecting drop size spectra using disdrometers and radar measurements in various rainfall events. They analyzed the data and compared actual rainfall rates with calculated rates obtained using the Z-DR method. The simulations and empirical analyses demonstrated that the Z-DR technique significantly reduced scatter in estimating rainfall rates compared to conventional methods. Furthermore, they explored the potential improvement by using empirical relations between Z-DR, ZH, and DSD parameters, especially the gamma distribution, which was more accurate in representing various DSD shapes.

\subsection{Backscattering}

Another experiment was performed in 1995  by Ali Tokay and Kenneth V. Beard \cite{tokay1996field,beard1987new,beard1991field}. They conducted a comprehensive field study to investigate the size spectra of oscillating raindrops during showers in Illinois at night. They aimed to determine the onset of raindrop oscillations, the causes behind these oscillations, and their effects on raindrop shapes. To capture raindrop streaks, they employed photographic measurements and backscattered light near the primary rainbow. The drop sizes were determined from fall speeds using strobe lights and continuously recorded with a JW disdrometer below the camera's sample volume.
Their remarkably designed experiments involved setting up spotlights and strobe lights to provide streak illumination at different angles for a wide range of raindrop sizes. The disdrometer sensor was positioned beneath the camera's sample volume to obtain drop size distributions. Raindrops scatter light in such a way that when they fall through the camera's sample volume, they produce bright streaks over a narrow range of angles corresponding to the primary rainbow. For non-oscillating raindrops, the rainbow angle gradually decreases as they fall through the light, resulting in a continuous streak with a specific color progression. However, for oscillating raindrops, their shape changes during the fall, causing variations in the rainbow angle. As a result, the streaks have interruptions or gaps due to changes in colors from red to yellow, green, or violet, depending on the amplitude of the oscillation. 

The rainbow streaks left behind by falling raindrops were recorded and compared with several models that considered pressure forces from turbulence and wind shear as the main cause of drop oscillations. It was found that drops oscillated invariably within the range of 1 mm to 4.2 mm. They concluded the most plausible cause of the drop oscillations was vortex shedding and aerodynamic fluctuations. The study showed that raindrop oscillations might drastically alter the polarisation of microwaves dispersed in rain, which has ramifications for radar reflectivity measurements and telecommunications networks.

\section{Surface Tension, Pressure, Young Laplace derivation}

\subsection{Surface Tension}

The surface tension of certain liquids against air at 1 atm and 25 ± C, expressed in millinewtons per meter, is provided in units. When the area of the interface is increased by a small increment, denoted as $dA$, the corresponding work required is equivalent to the surface energy contained within the additional surface area. 
\begin{equation}
    dW = \sigma dA   
\end{equation}
This relationship is analogous to the mechanical work performed against pressure during volume expansion. However, while volume expansion under positive pressure results in negative work, increasing the surface area requires positive work. This resistance to surface extension indicates the presence of a permanent internal tension known as surface tension, denoted as $\sigma$. Notably, surface tension can be defined as the force per unit length acting perpendicular to an imaginary line drawn on the interface. If we aim to stretch the interface along a straight line with a length of $L$ by a uniform amount $ds$, the new area $dA = Lds$, and the work required is given by $dW = \sigma Lds$. Consequently, the force acting orthogonally to the line is determined as $F = \sigma L$, or equivalently, $F/L = \sigma$. Hence, surface tension is essentially equivalent to the surface energy density, as reflected in their shared natural units of N/m and J/m$^2$.

\subsection{Pressure Excess in a sphere}
Consider a spherical liquid ball with a radius of $a$, created in the absence of any gravitational field. Surface tension exerts a contracting force on the ball, but this is counteracted by the development of an additional pressure, denoted as $\Delta p$, within the liquid. When we increase the radius by an increment of $da$, the work required against surface tension is given by 
\[
dW_1 = \sigma d(4\pi a^2) = \sigma 8\pi a da.
\]
This work is balanced by the thermodynamic work released by the expansion of volume and is expressed as
\[
dW_2 = - \Delta p (4\pi a^2) da.
\]
In equilibrium, the total work performed should be zero, leading to the relationship $dW_1 + dW_2 = 0$. From this, we can deduce that 
\[
\Delta p = \frac{2\sigma}{a}.
\]
Therefore, the pressure excess is inversely proportional to the radius of the sphere. 


\subsection{Laplace Young Derivation}

To derive the Laplace-Young equation from the equation $\Delta p = \sigma \left(\frac{dA}{dV}\right)$, we can start by considering a small volume element within a curved liquid interface. Let's denote the radius of curvature of the interface as $R$. The pressure difference $\Delta p$ across this curved interface can be expressed as $\Delta p = P_{\text{inside}} - P_{\text{outside}}$, where $P_{\text{inside}}$ and $P_{\text{outside}}$ represent the pressures inside and outside the curved interface, respectively.

Differntiating a small increase in surface area due to small increase in raidus
Differentiating a small increase in volume due to a small increase in radius

Since the change in radius is small (dr) any non linear terms of dr may be ignored


\section{Drop Distribution}
As the raindrop continues its descent, the air resistance continues to increase. This leads to The size distribution of raindrops, represented by the number of drops n(d) within a certain diameter range per unit volume, can be described using the equation $n(d) = n_0e^{(-d/<d>)}.$ Here, $n_0$ is a constant indicating the average spatial density of the drops, and `d' is an average diameter linked to the rate of rainfall (R). The steepness of the distribution is solely determined by the intensity of rainfall, with heavier storms exhibiting a wider range of drop sizes compared to lighter mists. Initially, the size distribution was derived from observations of raindrop impacts on absorbing paper and later confirmed through in-situ measurements conducted by aircraft flying through clouds and precipitation, as well as more precise measurements utilizing radar echo reflectivity.


Upon analyzing the experimental data, a general relation was found to accurately fit the observations. This relation is given by 
\begin{equation}
N(D) = N_0e^{-\frac{D}{\langle d \rangle}}
\end{equation}
where $\langle d \rangle = 41R^{-0.21} \text{cm}^{-1}$.

where D represents the diameter, N(D) represents the number of drops within the diameter range between D and D + SD per unit volume, and $N_0$ is the value of $N_0$  when D = 0. The value of 
$N_0$ is experimentally determined to be $0.08 cm^{-4}$ for any intensity of rainfall. Additionally, the relationship between the diameter and the rate of rainfall can be expressed as $<d> = 41R^{(-0.21)} cm^{-1}$ where R represents the rate of rainfall in millimetres per hour.

\begin{equation}
N(D) = N_0e^{-\frac{D}{\langle d \rangle}}
\end{equation}
where $\langle d \rangle = 41R^{-0.21} \text{cm}^{-1}$.

It is worth noting that for smaller diameters below approximately 1.5 mm, slight deviations and discrepancies are observed between the experimental data and the model predictions. These variations exist between the two sets of observations. Further research and analysis are needed to gain a deeper understanding of raindrop size dynamics and reconcile the discrepancies observed at smaller diameters. 

\subsection{Integral for rainfall intensity}

\begin{eqnarray}
    R = \frac{n_0}{6 \sqrt{g}} \left(\frac{\rho}{\rho_a}\right)^{\frac{3}{2}} d_0^2 \int x^{\frac{7}{2}} p(x) dx \\
    \frac{d_0}{R} \sim R^{-\frac{2}{9}} 
\end{eqnarray}

Equation (5) provides a precise estimate of the scaling exponent in equation (2), which was originally measured by Marshall and Palmer to be approximately $0.21$. The pre-factor in equation (2) can be quantitatively determined once the mechanism governing the drop size distribution $P(d)$ is fully understood.

\begin{equation}
\frac{{\ddot{z}}}{{g}} = -1 + \frac{{CD}}{{2}} \cdot \frac{{\rho_a}}{{\rho}} \cdot \frac{{U^2}}{{g}} h
\end{equation}

The vertical altitude $z(t)$ of a liquid drop of volume $\Omega = \frac{\pi d^3}{6}$ and density $\rho$ falling under gravity $g$ in an ascending air stream of density $\rho_a$ and velocity $V$ is governed by

where $U = -\dot{z} + V$ is the relative velocity between the drop and the air. 
\\ rough pancake $\Omega = \frac{{\pi R^2 h}}{6}$. 
\\ $V = 0$, $U = -\dot{z}(d_0) \approx \sqrt{\frac{{\rho}}{{\rho_a}} g d_0}$ free-fall or terminal velocity.

To derive the equation for terminal velocity, we start with the equation governing the vertical altitude (z) of a falling liquid drop:

\[ \frac{z}{g} = -1 + \frac{CD}{2} \cdot \frac{\rho_a}{\rho} \cdot \frac{U^2}{g}h \]

where:
- \( g \) is the acceleration due to gravity
- \( CD \) is the drag coefficient (assumed to be of order unity, so \( CD = 2 \))
- \( \rho_a \) is the density of the air
- \( \rho \) is the density of the liquid drop
- \( U = -\dot{z} + V \) is the relative velocity between the drop and the air, where \( V \) is the velocity of the air
- \( h \) is the thickness of the drop in its pancake shape
- \( \Omega = \pi R^2 h \) is the volume of the drop, where \( R \) is the radius of the pancake shape

Since we are interested in the terminal velocity, we assume that at terminal velocity, the acceleration (\( \ddot{z} \)) is zero, i.e., \( \ddot{z} = 0 \). Therefore, we have \( \dot{z} = 0 \) at terminal velocity.

Substituting \( \dot{z} = 0 \) and \( U = \text{terminal velocity} \) into the equation, we get:

\[ \frac{z}{g} = -1 + \frac{CD}{2} \cdot \frac{\rho_a}{\rho} \cdot \frac{(\text{terminal velocity})^2}{g} h \]

Solving for the terminal velocity, we have:

\[ (\text{terminal velocity})^2 = \frac{2}{CD} \cdot \frac{\rho}{\rho_a} \cdot g \cdot \frac{z}{h} \]

Now, since \( h = d_0 \) and \(C_d=2)\) as it along the order unity, we can rewrite the equation as:

\[ (\text{terminal velocity})^2 = \frac{2}{2} \cdot \frac{\rho}{\rho_a} \cdot g \cdot \frac{z}{d_0} \]

Finally, taking the square root of both sides, we get the equation for terminal velocity:

\[ \text{terminal velocity} = \sqrt{\cdot \frac{\rho}{\rho_a} \cdot g \cdot \frac{z}{d_0}} \]


where $U = \sqrt{\frac{{\rho}}{{\rho_a}}} g d$ represents the free-fall velocity of a drop of size $d$.In the derivation 2 assumptions were Cd was along the order of unity and $h=d_0$
so that this equation roughly yield the order of the terminal velocity

\begin{equation}
R = \int n(d) \cdot \frac{{\pi d^3}}{{6}} \cdot U \, dd
\end{equation}

\section{Drop formation}
It is important to clarify that the distinct teardrop shape is not a shape ever assumed by raindrops, but rather appears when a drop of water detaches from a surface. Raindrops, on the other hand, undergo a complex life cycle.

The life cycle of a raindrop begins within clouds, where high altitudes and low temperatures cool water vapour below its condensation point. These droplets condense and coalesce around nucleation points, which are tiny particles of dust or soil suspended in the clouds. These nucleation sites are essential for raindrop formation and can include various substances such as dust, dirt, pollen, plant fragments, mushroom spores, and even bacteria. The presence of nucleation sites serves as a physical requirement for raindrop formation and is the working principle in several cloud-seeding techniques.

Without nucleation sites, raindrops would never reach the critical size necessary for the equilibrium between the free energy required to form a surface and the free energy released during volume formation. Prior to reaching this critical size, raindrops are more likely to decrease in size rather than grow as the free energy required to increase surface area is greater than what would be released by increasing the volume. The process of raindrop formation involves the aggregation and collision of water droplets around nucleation sites. As the droplets collide and merge, they gradually grow in size, accumulating additional water molecules. This growth continues until the raindrop reaches a size where it can no longer be suspended in the cloud and falls freely towards the surface of the earth.

\[
\Delta p = \frac{2\sigma}{a}.
\]
Therefore, the pressure excess is inversely proportional to the radius of the sphere. 

\section{Burst Cycle and Fragmentation}
During its descent, a raindrop undergoes a series of shape transformations due to the interplay between surface tension and air resistance. Initially, the raindrop adopts a spherical configuration in order to minimize surface tension forces. However, as it accelerates towards the Earth, the force of air resistance becomes increasingly significant. As this force acts primarily on one side it causes the raindrop to partially flatten assuming a hemispherical shape 

\begin{figure}[htp]
    \centering
    \includegraphics[width=4cm]{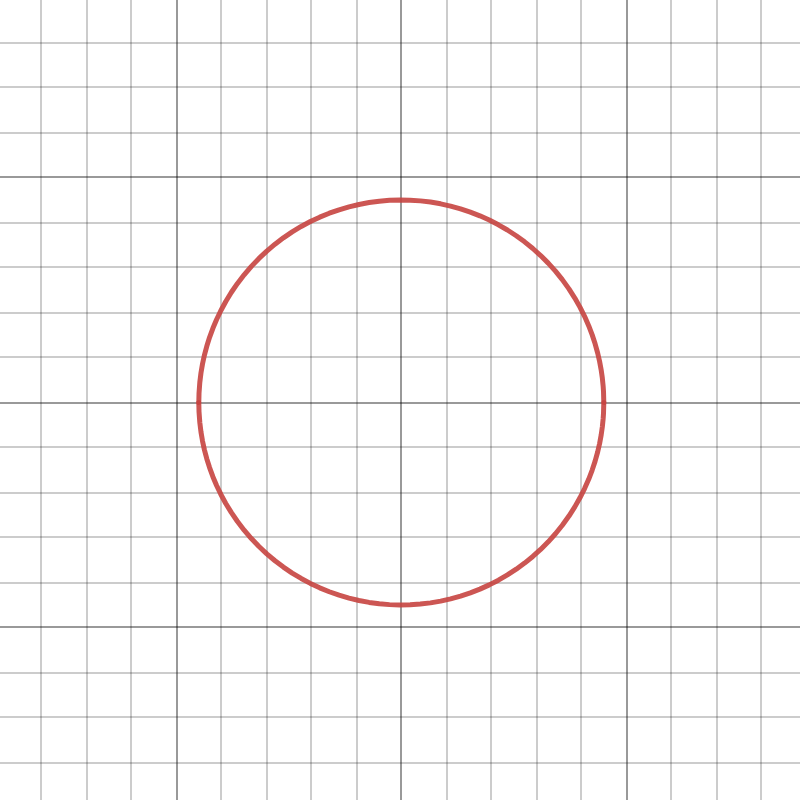}
     \includegraphics[width=4cm]{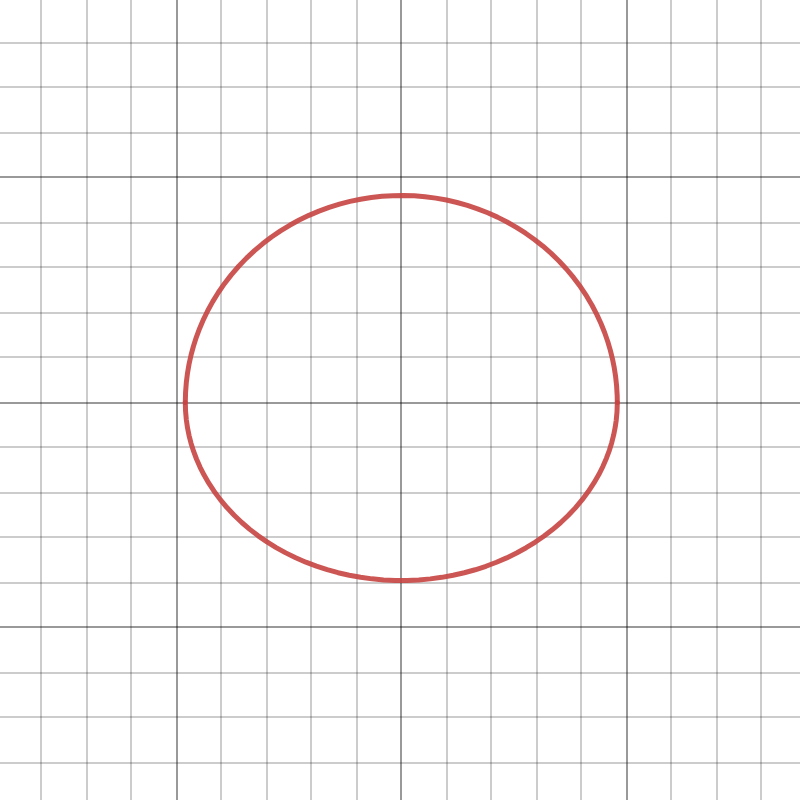}
      \includegraphics[width=4cm]{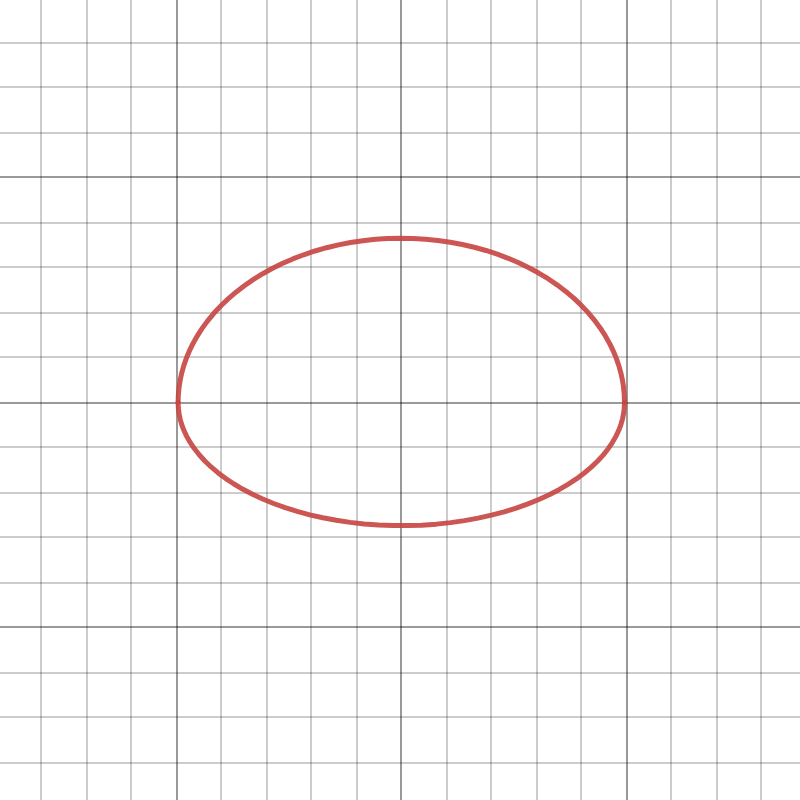}
       \includegraphics[width=4cm]{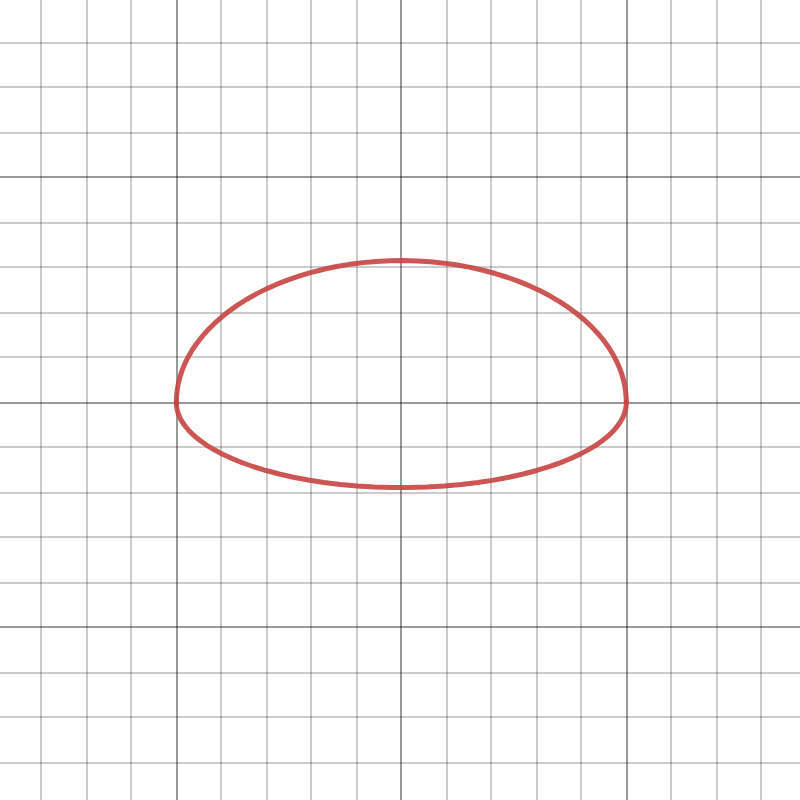}
    \caption{Asymmetrical axial flattening of raindrop}
    \label{fig:burst cycle}
\end{figure}

As the raindrop continues its descent, the air resistance continues to increase. This leads to further deformation of the raindrop, causing the flattened side to become progressively more concave, in a stage known as Inflation. The raindrop now resembles a parachute, with a toroidal rim around its edge. However, the deformation of the raindrop is not without limits. As the air resistance reaches a critical point, the toroidal rim becomes unstablen in a stage called destabilistion. At this stage, the raindrop undergoes a phenomenon known as ``bag break-up." The instability of the rim causes the raindrop to burst into smaller fragments, which are distributed in various sizes in a process known as fragmentation.

\begin{figure}[htp]
    \centering
       \includegraphics[width=4cm]{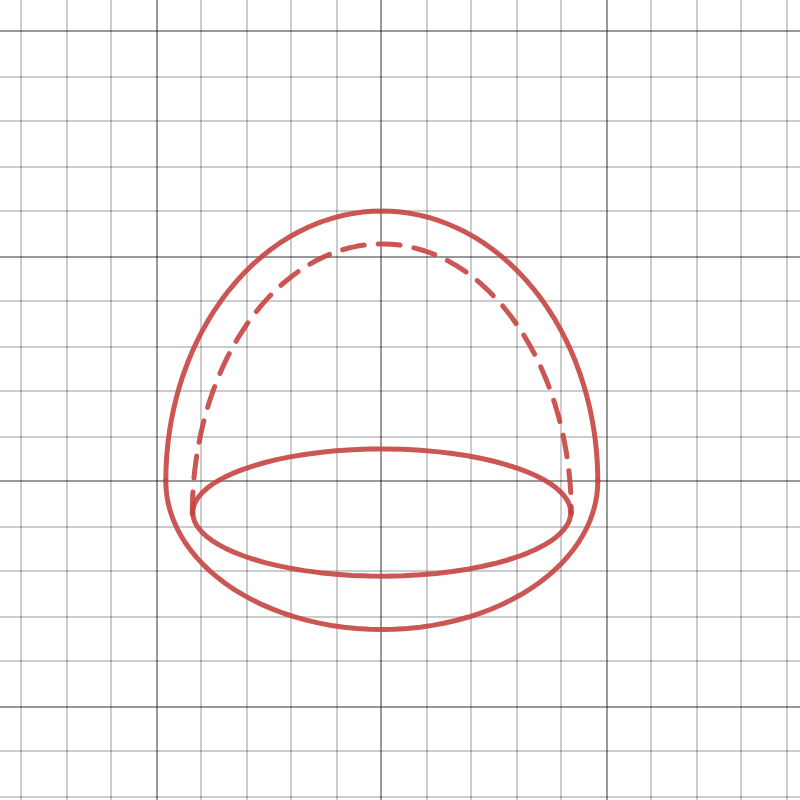}
    \caption{T burst}
    \label{fig:burst cycle}
\end{figure}






\section{Future Investigation}
The current research could be extended by investigating how environmental factors such as pollution, altitude, humidity can affect raindrop distribution and rainfall intensity. 

Furthermore classification algorithms could be trained on current data sets related to raindrop formation. This could then be used to predict rainfall patterns and potentially identify new correlations and contribute to a more comprehensive understanding of factors that affect rainfall. These algorithms could possibly improve current models and make better adjustments for atmospheric attenuation that impedes the propagation of radio wave signals. 

A greater focus could be shed on the microphysical processes that occur in clouds and lead to sucessful raindrop nucleation. This research could then  be used to inform and improve the effectivness of methods such as cloud seeding which operate primarily by on increasing the number of nucleation sites for drop condensation. 

\section{Conclusion}
In this work, we reviewed a few methods to understand raindrop formation and how they can be measured experimentally. We studied theory and principles governing raindrop formation, delving into the mathematical parameters associated with the natural occurrence of this phenomenon. Understanding the dynamics of how raindrops take shape is fundamental, so we focused on this aspect extensively.  We explored drop distribution in various rainfall intensities and the topological transforms in larger droplets as they fall. Additionally, we delved into studying how raindrop distribution varies across different levels of rainfall intensity. This variation is vital in comprehending the diversity of raindrop sizes we observe during various types of rain. Furthermore, we paid special attention to the transformations larger raindrops undergo as they descend through the atmosphere.Ultimately, our work aims to uncover the subtle details of raindrop formation, emphasizing the importance of such understanding in the broader field of meteorology and atmospheric science.

\bibliographystyle{unsrt}
\bibliography{sample.bib}


\end{document}